\documentclass[12pt]{iopart}

\expandafter\let\csname equation*\endcsname\relax

\expandafter\let\csname endequation*\endcsname\relax

\usepackage[utf8]{inputenc}
\usepackage{amssymb,amsfonts,dsfont,xspace,graphicx,relsize,bm,mathtools,amsthm,soul}

\usepackage{hyperref}
\usepackage[monochrome]{xcolor}

\usepackage[margin=1in]{geometry}
\hypersetup{
	colorlinks,
	linkcolor={red!100!black},
	citecolor={blue!100!black},
}

\usepackage{subcaption}
\usepackage{bbm}
\usepackage{mleftright}
\usepackage{hyperref}
\usepackage{tipa}
\usepackage{ulem}
\usepackage{enumitem}
\usepackage{bm}
\usepackage{bbold}
\usepackage{physics}

\newcommand{\cE}{{\mathcal{E}}}

\newcommand{\X}{{\mathsf{X}}}

\makeatletter
\let\X\@undefined
\makeatother
\newcommand{\X}{{\mathcal{X}}}

\theoremstyle{plain}

\usepackage{algorithm}
\usepackage{algpseudocode}

\begin{document}

\title{Probabilistic pulse-position modulation for classical communication on quantum channels}

\author{Farzad Kianvash, Matteo Rosati}

\address{Dipartimento di Ingegneria Civile, Informatica e delle Tecnologie Aeronautiche, Università Roma Tre, Via Vito Volterra 62, I-00146 Rome, Italy.}
\ead{farzad.kianvash@uniroma3.it}
\vspace{10pt}
\begin{indented}
\item[]\date{}
\end{indented}

\begin{abstract}
Classical communication over lossy quantum channels is an essential topic in quantum information theory, with practical implications for optical-fiber and free-space communications. Multi-phase Hadamard codes, based on coherent-state Binary Phase-Shift Keying (BPSK) modulation and decoded using vacuum-or-pulse (VP) detectors, offer a promising approach for achieving high communication rates while relying only on linear optics and single-photon detectors (SPDs). However, their performance does not reach the ultimate Holevo limit. In this work, we propose a generalization of Hadamard codes that distributes the signal across multiple modes with optimized probabilities, rather than concentrating it in a single mode, dubbed probabilistic pulse-position modulation (PPPM). 
We derive an achievable communication rate, demonstrating that our PPPM can outperform traditional Hadamard codes in certain intermediate energy regimes. 
\end{abstract}

%
%
%
%
%


\section{Introduction}
Quantum communication (QC) explores the ultimate limits of information transmission as dictated by quantum mechanics. One of the tasks made possible by QC is the transmission of bits using quantum states of light at a rate surpassing the Shannon capacity and achieving the Holevo capacity—the theoretical maximum for transmitting classical information through quantum channels~\cite{Banaszek2020,HolevoBOOK}. Achieving this capacity requires encoding classical information into quantum states, transmitting them over multiple uses of the quantum channel, and employing an optimal decoding strategy. This process may involve encoding strategies that utilize entanglement~\cite{serafini2023quantum} and decoding methods that employ joint-detection receivers (JDRs)~\cite{hausladen1994pretty,schumacher1997sending,lloyd2011sequential,giovannettiachieving,sen2012achieving,wilde2013towards,wilde2012polar,Cui2023,Rosati2024b,Rosati2025}. JDRs perform collective quantum measurements on sequences of signals, enabling capacities that surpass those achievable with single-symbol detection strategies.  However, the practical realization of JDRs
remains challenging, constrained by technological limitations in manipulating
continuous-variable quantum states of light.\\

In the specific case of lossy channels, which accurately model realistic noise in optical-fiber and free-space transmission, the authors in~\cite{giovannetti2004classical} demonstrated that entangled states are not necessary in the encoding process to achieve the Holevo capacity. Instead, the Holevo quantity can be maximized using a sequence of coherent states, a method that is relatively straightforward to implement experimentally. On the other hand, to the best of our knowledge there is no known optimal decoder that relies solely on linear optics and single-photon detectors (SPDs). For this reason, extensive research~\cite{Klimek2015,Klimek2015a,Rosati2021,Rosati2025a,Rosati2025,Rosati16c,Rengaswamy2020,Cui2023} has been conducted to study coherent-state encodings with receivers that use only linear optics and SPDs, given their  relevance in practical implementations.

In particular, in binary phase-shift keying (BPSK) each codeword corresponds to a sequence of coherent states $\ket{\alpha_1\alpha_2...\alpha_N}$, where $\alpha_i$ is either $\alpha$ or $-\alpha$, and $\alpha$ is a complex number.  BPSK quantum codes are particularly useful in low-photon-number regimes i.e. $\abs{\alpha}<<1$, as their performance is close to the Holevo limit.\\
The code introduce in~\cite{Guha2011a}, often referred to as the two-phase Hadamard code, is designed to discriminate a subset of BPSK-modulated sequences with SPDs. In this scheme, \(2N\) classical messages (\(N = 2^n\), where \(n\) is a natural number) are encoded into a subset of \(N\)-mode BPSK coherent states, whose first moments are orthogonal to each other, and including both these states and their oppositely signed counterparts.

It is well-known that, using a series of \(50\)-\(50\) beam splitters, these states can be transformed into configurations where the entire signal is concentrated in a single mode with either a positive or negative phase, while all other modes remain in the vacuum state~\cite{Guha11,Rosati16c,Rosati2025a}. Specifically, the final states take the form of a pulse-position modulation (PPM)~\cite{Essiambre2023}:
\[
\ket{0}^{\otimes(i-1)} \otimes \ket{\pm \sqrt{N}\alpha} \otimes \ket{0}^{\otimes(N-i)}, \quad 0 < i \leq N.
\]

For decoding, Vacuum or Pulse (VP) detectors are employed~\cite{Guha2011a,Rosati2016}. These first identify the mode containing the signal and then determine its phase using a Dolinar detector~\cite{Dolinar1973,Cook2007,Becerra2013a}. Although the performance of Hadamard codes does not reach the Holevo limit, in the low-energy regime (\(\abs{\alpha} <.01\)), they outperform all other codes that rely solely on SPDs. 
 \\
In this paper, we propose a generalization of Hadamard codes that can be transformed via linear optics into a probabilistic pulse-position modulation (PPPM). In this new code, the messages are not only encoded into states where the signal can be concentrated into a single mode, but they also have a probability $1-p$ of being encoded into states where the signal can be concentrated in two or more modes. Naturally, in the limit $p=1$ one can recover the Hadamard code, whereas in the limit $p=0$ one can obtain multi-pulse codes similar to those studied in~\cite{Rosati2025a}. However, it turns out that one can optimize such probability to get an advantage over previous codes in the literature. We show that there exists an intermediate energy regime where PPPM codes present a moderate advantage over Hadamard codes as well as single-mode strategies.

\section{Preliminaries}
In this section, we introduce several key concepts essential for presenting our main result. First, we provide a brief overview of the Dolinar receiver~\cite{Dolinar1973,Holevo1982,Takeoka2005,Assalini2011,Sych2014,Cook2007,Geremia2004}, and we discuss the vacuum-or-pulse detector (VP-detector). Finally, we explore Hadamard codes for lossy channels. We stress that all the protocols discussed in this article can be realized only by beam-splitters, displacements and SPDs. A beam-splitter with parameter $ 0\leq\eta\leq 1$, $\hat{B}_\eta$ acts as follows on two coherent states
\begin{equation}
    \nonumber \hat{B}_\eta\ket{\alpha}\ket{\beta}=|\sqrt{\eta}\alpha + \sqrt{1-\eta}\beta\rangle  |-\sqrt{1-\eta}\alpha + \sqrt{\eta}\beta\rangle\, .
\end{equation}
The displacement operator shifts any coherent state by a fixed amount i.e. $\hat{D}(\alpha)\ket{\beta}=\ket{\alpha +\beta}$.

\subsection{Dolinar Receiver and VP-Detector}
The Dolinar receiver~\cite{Dolinar1973} is a quantum-optimal detection strategy designed to distinguish between two non-orthogonal coherent states, $|\alpha\rangle$ and $|-\alpha\rangle$, with the minimum possible probability of error, achieving the Helstrom limit~\cite{Helstrom1976} for the probability of error i.e., $P_{\text{Hel}}(\mathcal{E})=\frac{1}{2}(1-\sqrt{1-e^{-4\mathcal{E}}})$ where $\mathcal{E}=\abs{\alpha}^2$. It operates using an adaptive feedback mechanism that dynamically adjusts a displacement operation based on previous photon detection results. The input state is  plit into $N$ smaller-amplitude beams i.e. $\ket{\frac{\pm \alpha}{\sqrt{N}}}^{\otimes N}$, each processed sequentially. The receiver applies a displacement $\hat{D}(\frac{\alpha}{N})$ or $\hat{D}(\frac{-\alpha}{N})$ to one of the split states before detection, steering the state closer to vacuum if the hypothesis is correct. A SPD  measures the outcome of each beam, and the feedback switches the displacement if photons are detected, indicating the current hypothesis is likely incorrect. This iterative process refines the state hypothesis over time, with the sequence of detection results providing increasingly accurate information about the input state. \\

Imagine a situation where it is uncertain whether a pulse exists in a given mode. If a pulse is present, its state is either $\ket{\alpha}$ or $\ket{-\alpha}$. A vacuum-or-pulse detector (VP) detector~\cite{Guha2011a,Rosati2016} is designed to distinguish between these scenarios. \

The VP-detector operates as follows. First, a beam splitter with small transmissivity divides the received state into two parts. Let us assume that the smaller fraction has an amplitude of $\frac{1}{\sqrt{T}}$ of the original state. This smaller beam is then measured using a photodetector. If the photodetector clicks, we can conclude that the original state is either $\ket{\alpha}$ or $\ket{-\alpha}$. The remaining beam, with a larger amplitude, is subsequently fed into a Dolinar receiver to distinguish between these states.

If the photodetector does not click, the process is repeated using the beam with the larger amplitude until a click is observed. This involves passing the beam through a beam splitter again and performing another measurement with the photodetector. Note that the transmissivity parameter of the beam splitter is adjusted in each round such that the smaller beam always has an amplitude of $\frac{1}{\sqrt{T}}$ of the original state. If the photodetector clicks during any of these repetitions, the rest of the beam is sent to a Dolinar receiver. Conversely, if no clicks are observed after $T$ steps, we conclude that the input state is vacuum.\\ 

Taking the limit of large $T$ one can show that~\cite{Guha2011a,Rosati2016} the probability of guessing $\ket{\alpha}$ if the input state is $\ket{\alpha}$ or $\ket{-\alpha}$ are respectively as follows
\begin{align}\label{pvp prob}
P_{VP}(\alpha|\alpha)&=\int_{e^\mathcal{-E}}^1\frac{1}{2}(1+\sqrt{1-e^{-4\mathcal{E}}/t^4})dt\, ,\\
P_{VP}(-\alpha|\alpha)&=\int_{e^\mathcal{-E}}^1\frac{1}{2}(1-\sqrt{1-e^{-4\mathcal{E}}/t^4})dt\, ,
\end{align}
where $\mathcal{E}=\abs{\alpha}^2$ is the energy of the input state in Planck units. Note that it is straightforward to show that $P_{VP}(-\alpha|-\alpha)=P_{VP}(\alpha|\alpha)$ and $P_{VP}(\alpha|-\alpha)=P_{VP}(-\alpha|\alpha)$. In addition, if the input state is vacuum one will always guess vacuum. while if the input state is $\ket{\pm \alpha}$ the probability of guessing vacuum is equal to the probability of not clicking 
\begin{equation}
    P_{VP}(0|\alpha)=P_{VP}(0|-\alpha)=\abs{\bra{0}\ket{\alpha}}^2=
    e^{-\mathcal{E}}
\end{equation}

\subsection{Hadamard Codes with BPSK Modulation}

In this section, we review Hadamard codes~\cite{Guha2011} with BPSK modulation. A Hadamard matrix is a square matrix of size \( N \times N \), where each entry is either \( +1 \) or \( -1 \), and all rows (and columns) are mutually orthogonal. This implies that the matrix \( H \) satisfies the property \( H H^\top = N I \), where \( H^\top \) is the transpose of \( H \), and \( I \) is the identity matrix of size \( N \).

A Hadamard matrix of size \( N \) can be constructed recursively using the formula:
\[
H_{2N} =
\begin{bmatrix}
H_N & H_N \\
H_N & -H_N
\end{bmatrix},
\]
starting with the base case \( H_1 = [1] \). For instance, the Hadamard matrix of size 4 is:
\[
H_4 =
\begin{bmatrix}
1 &  1 &  1 &  1 \\
1 & -1 &  1 & -1 \\
1 &  1 & -1 & -1 \\
1 & -1 & -1 &  1
\end{bmatrix}.
\]

A Hadamard code of size \( N \) with BPSK modulation maps \( 2N \) messages to coherent states across \( N \) modes, represented as \( \ket{\alpha_1, \alpha_2, \dots, \alpha_N} \), where \( \alpha_i \) is either \( \alpha \) or \( -\alpha \). The signs of the coherent states are determined by the rows of the Hadamard matrix. Additionally, for any state in the code space, the state with the opposite phase is also part of the code-space.

For example, the code space for \( N=4 \) is:
\begin{equation}
\begin{aligned}
\ket{\psi_1} &= \ket{\alpha, \alpha, \alpha, \alpha}, & \quad \ket{\bar{\psi}_1} &= \ket{-\alpha, -\alpha, -\alpha, -\alpha}, \\
\ket{\psi_2} &= \ket{\alpha, -\alpha, \alpha, -\alpha}, & \quad \ket{\bar{\psi}_2} &= \ket{-\alpha, \alpha, -\alpha, \alpha}, \\
\ket{\psi_3} &= \ket{\alpha, \alpha, -\alpha, -\alpha}, & \quad \ket{\bar{\psi}_3} &= \ket{-\alpha, -\alpha, \alpha, \alpha}, \\
\ket{\psi_4} &= \ket{\alpha, -\alpha, -\alpha, \alpha}, & \quad \ket{\bar{\psi}_4} &= \ket{-\alpha, \alpha, \alpha, -\alpha}.
\end{aligned}
\end{equation}

Since the rows of the Hadamard matrix are orthogonal, the states in the code space can be transformed into a basis where the signal is concentrated in a single mode using a sequence of 50-50 beam splitters \( \hat{B}_{\frac{1}{2}} \).

For \( N = 4 \), applying the following unitary transformation achieves this mapping:
\[
\hat{U}_4 = \hat{B}^{24}_{\frac{1}{2}} \hat{B}^{13}_{\frac{1}{2}} \hat{B}^{34}_{\frac{1}{2}} \hat{B}^{12}_{\frac{1}{2}},
\]
where \( \hat{B}^{ij}_{\frac{1}{2}} \) represents a beam-splitter between mode \( i \) and mode \( j \). After applying \( \hat{U}_4 \), the code-space states transform into a pulse-position modulation (PPM) with additional phase-modulation on each pulse:
\begin{equation}\label{codespace 1}
\begin{aligned}
\hat{U}_4 \ket{\psi_1} &= \ket{\sqrt{4}\alpha, 0, 0, 0}, & \quad \hat{U}_4 \ket{\bar{\psi}_1} &= \ket{-\sqrt{4}\alpha, 0, 0, 0}, \\
\hat{U}_4 \ket{\psi_2} &= \ket{0, \sqrt{4}\alpha, 0, 0}, & \quad \hat{U}_4 \ket{\bar{\psi}_2} &= \ket{0, -\sqrt{4}\alpha, 0, 0}, \\
\hat{U}_4 \ket{\psi_3} &= \ket{0, 0, \sqrt{4}\alpha, 0}, & \quad \hat{U}_4 \ket{\bar{\psi}_3} &= \ket{0, 0, -\sqrt{4}\alpha, 0}, \\
\hat{U}_4 \ket{\psi_4} &= \ket{0, 0, 0, \sqrt{4}\alpha}, & \quad \hat{U}_4 \ket{\bar{\psi}_4} &= \ket{0, 0, 0, -\sqrt{4}\alpha}.
\end{aligned}
\end{equation}
We define states $\ket{k,s,\mathcal{E},N}$ as an $N$-mode state where the signal is concentrated in mode $k$ and the average energy is $\mathcal{E}$, while $s=\pm$ determines the phase. For instance,$\ket{2,-,\mathcal{E},4}=\ket{0,-\sqrt{4E},0,0}$. 

\begin{figure}[h!]
    \centering
    \includegraphics[width=0.8\textwidth]{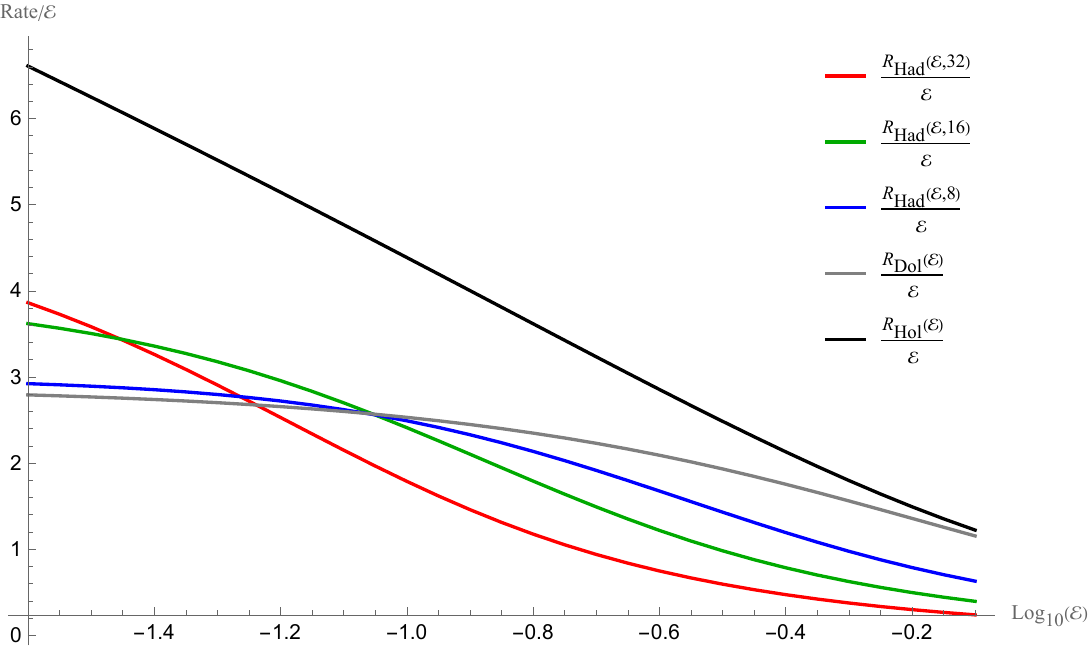}
    \caption{The comparison between different rates for the classical capacity of lossy channel with BPSK modulation. The Holevo rate $R_{\text{Hol}}$ in Eq.~\ref{Hol} is the ultimate rate, and $R_{\text{Dol}}$ in Eq.~\ref{Dol}, and the Hadamard rates $R_{\text{Hol}}$ in Eq.~\ref{Had rate}}
    \label{fig:plot1}
\end{figure}

After carrying out this transformation, the message can be decoded using a VP detector. The rate of transmission of information is given by the mutual information, normalized  by the number of channel uses $N$. To find the mutual information we need to calculate the conditional probability distribution $P^{(\mathcal{E},N)}_{\text{Had}}(Y|X)$ where $x=(k_x,s_x)$, $y=(k_y,s_y)$:
\begin{equation}
    P^{(\mathcal{E},N)}_{\text{Had}}(Y|X)= 
\begin{cases} 
P_{VP}(s_y\sqrt{N\mathcal{E}}|s_x\sqrt{N\mathcal{E}}) \delta_{k_y,k_x},  & \text{if } y\neq\text{Error}\\
(1-e^{-N\mathcal{E}}), & \text{otherwise},
\end{cases}
\end{equation}
Where $y=$Error corresponds to the outcome where none of the detectors click.
Doing some algebra one can show that the rate is as follows
\begin{align} \label{Had rate}
   R_{\text{}Had}(\mathcal{E},N)=2 \left(h_1\left[\frac{1-e^{-N\mathcal{E}} }{2 N}\right]-\frac{h_1[P_{VP}(\sqrt{N\mathcal{E}}|\sqrt{N\mathcal{E}})]+h1[P_{VP}(\sqrt{N\mathcal{E}}|-\sqrt{N\mathcal{E}})]}{2 N}\right)\,
\end{align}
where $h_1[x]=-x\log_2[x]$. One can compare this rate with that obtained by simply distinguishing between $\ket{\alpha}$ and $\ket{-\alpha}$ using a Dolinar receiver on each mode\cite{Rosati2023}: 
\begin{equation}\label{Dol}
    R_{\text{Dol}}(\mathcal{E})=1-h_2[P_{\text{Hel}}(\mathcal{E})]\, ,
\end{equation}
where $h_2[x]=h_1[x]+h_1[1-x]$. Dolinar rate is the maximum achievable rate using just a single mode.
The maximum achievable rate with a BPSK modulation is given by the Holevo capacity for alphabet $\{\ket{\pm\sqrt{\cE}}\}$, which relies on optimal JDRs whose practical implementation is not known to date:
\begin{equation}\label{Hol}
    R_{\text{Hol}}(\mathcal{E})=h_2\left[\frac{1-e^{-2\mathcal{E}}}{2}\right]
\end{equation}
As one can note in Fig.~\ref{fig:plot1}, in the low energy regime, the performance of Hadamard codes gets better increasing $N$, and for large enough $N$ the Hadamard code outperforms Dolinar in the low-energy regime.

\section{Probabilistic pulse-position modulation code}

Hadamard codes encode classical messages into states where the signal is concentrated in a single mode, i.e., equivalent to a PPM code up to a unitary transformation. In this work, we consider a generalization of thereof: with probability $p$, messages are encoded into states where the energy is concentrated in one mode, and with probability $1-p$, they are encoded into states where the signal is distributed across two modes, while maintaining a constant average energy. Setting $p = 1$ reduces this scheme to the standard Hadamard code, while for $p=0$ one can obtain a two-pulse-convertible code similar to those studied in~\cite{Rosati2025a}, hence we dub our code probabilistic pulse-position modulation (PPPM). This generalization increases the code space but also introduces greater complexity in the decoding process. 

To make the presentation clear, we define the notation $\ket{(k_1,k_2),(s_1,s_2),\mathcal{E},N}$, which represents an $N$-mode state where the signal is concentrated on modes $k_1$ and $k_2$, with phases $s_1$ and $s_2$, respectively. For example, $\ket{(1,3),(-,+),\mathcal{E},4} = \ket{-\sqrt{2\mathcal{E}}, 0, \sqrt{2\mathcal{E}}, 0}$. A straightforward calculation reveals that, for \( N \) modes, this code can encode \(N_{\rm tot} = 4\binom{N}{2} + 2N \) distinct messages. This is greater than the number of messages the normal Hadamard code can encode, which is \( 2N \). \\
\subsection{Decoding Procedure}
The decoding process begins with running VP detectors simultaneously on all modes. Suppose the first click occurs at time $t_1$ in the VP detector of mode $i$. At this point, the VP detector for mode $i$ is frozen, while the others continue. If no additional clicks occur, the state is assumed to be of the form $\ket{i,s,\mathcal{E},N}$, where the signal is concentrated only in mode $i$. A Dolinar detector is then used to determine the phase $s$ of the state on mode $i$.

Instead, if a second click occurs at time $t_2 > t_1$ in mode $j\neq i$, the state is identified as $\ket{(i,j),(s_1,s_2),\mathcal{E},N}$, where the signal is distributed across modes $i$ and $j$. The remaining state of the populated modes then is $\ket{s_1\sqrt{\frac{T-t_1}{T}\mathcal{E}}}_i\otimes\ket{s_2\sqrt{\frac{T-t_2}{T}\mathcal{E}}}_j$. Therefore, mode $i$ is mapped to $\ket{s_1\sqrt{\frac{T-t_2}{T}\mathcal{E}}}_i$ by blocking part of the signal, which is always possible because $t_2 > t_1$.

A 50-50 beamsplitter is subsequently applied between modes $i$ and $j$, mapping the state to one where the signal is concentrated in either mode $i$ or mode $j$. Finally, another VP detector is used to determine the phases of the signal in the respective modes. If the VP detector does not click in this last stage, we will guess an Error of type $1$ has occured.\\
Finally, if we do not get any click at all, we classify the outcome as an Error of type 2. Separating these two error types is fundamental because in Error type 1 we retain information about the position of the signals, whereas in Error type 2 we lack any information about the received state.

\subsection{Computing the Rate}
To compute the rate, we need to identify the input-output conditional probability distribution. For clarity, we introduce a concentration parameter $c$, which can take values of one or two. This parameter indicates whether we are in the code-space with states concentrated in one mode ($c = 1$) or two modes ($c = 2$). Once again we need to write the conditional entropy $P^{(\mathcal{E},N)}_{PPPM}(Y|X)$ where $x$ can take values $x=((k_{x_1},k_{x_2}),(s_{x_1},s_{x_2}),c_x=2)$ or $x=((k_{x}),(s_{x}),c_x=1)$ and similarly for $y$. Then we can write

\begin{align}
    P^{(\mathcal{E},N)}_{PPPM}(Y|X)=\begin{cases} 
P_{VP}(s_y\sqrt{N\mathcal{E}}|s_x\sqrt{N\mathcal{E}}) \delta_{k_y,k_x}, & \text{if } c_x=c_y=1,\\
\frac{(1-e^{\frac{-N\mathcal{E}}{2}})e^{\frac{-N\mathcal{E}}{2}}}{2}\delta_{k_{1x},k_y}, & \text{if } c_x=2,\, c_y=1,\\
0, & \text{if } c_x=1,\, c_y=2,\\
(P_{S}(\mathcal{E},N)\delta_{(s_{x_1},s_{x_2}),(s_{y_1},s_{y_2})})+\\P_{E}(\mathcal{E},N)(1-\delta_{(s_{x_1},s_{x_2}),(s_{y_1},s_{y_2})}))\delta_{(k_{x_1},k_{x_2}),(k_{y_1},k_{y_2})}, & \text{if } c_x=2,\, c_y=2,\\
{\color{red}\frac12} P_{nc}(\mathcal{E},N)\delta_{(k_{x_1},k_{x_2}),(k_{y_1},k_{y_2})}, & \text{if } y=\text{Error 1},\\
{\color{red}\frac1{N_{\rm tot}}} (1-e^{-N\mathcal{E}}), & \text{if } y=\text{Error 2},\\
\end{cases}
\end{align}

When $c_x = c_y = 1$, the scenario is similar to the Hadamard code, and the conditional entropy is the same. For the case where $c_x = 2$ and $c_y = 1$, the received state has two modes with signals; one clicks, and the other does not. The probability of this occurrence is $(1 - e^{-\frac{N\mathcal{E}}{2}})e^{-\frac{N\mathcal{E}}{2}}$, and the factor $\frac{1}{2}$ accounts for the fact that half of the time, the phase is guessed correctly. Note that this is only an approximation of the true strategy and provides a lower bound for the achievable mutual information.

The case where $c_x = 1$ and $c_y = 2$ does not occur because when the received state has one mode with a signal, we can never observe two clicks in different modes, which simplifies the analysis for this scenario.

The last case where $c_x=c_y=2$ is when we receive a state where the signal exists in two modes and both click. In this case, we will not make any mistakes in guessing $k_{y_1}$ and $k_{y_2}$, however we can still mistake the phases. The probability of guessing correctly (incorrectly) the phase is $P_S$  ($P_E$) and given by the following expression:

\begin{align}
P_S(\mathcal{E},N)&=\lim_{T\rightarrow\infty} \sum^{T}_{t_1=1}\sum_{t_2=t_1+1}^{T}2(1-e^{\frac{-N\mathcal{E}/2}{T}})^2e^{\frac{-(t_1+t_2-2)N\mathcal{E}/2}{T}}P_{VP}\left(\sqrt{\frac{N\mathcal{E}(T-t_2)}{T}}\Bigg|\sqrt{\frac{N\mathcal{E}(T-t_2)}{T}}\right)\\
P_E(\mathcal{E},N)&=\lim_{T\rightarrow\infty} \sum^{T}_{t_1=1}\sum_{t_2=t_1+1}^{T}2(1-e^{\frac{-N\mathcal{E}/2}{T}})^2e^{\frac{-(t_1+t_2-2)N\mathcal{E}/2}{T}}P_{VP}\left(-\sqrt{\frac{N\mathcal{E}(T-t_2)}{T}}\Bigg|\sqrt{\frac{N\mathcal{E}(T-t_2)}{T}}\right)
\end{align}
We explain the derivation of the first \( P_S(\mathcal{E}, N) \), which represents the probability of obtaining two clicks when the given state satisfies \( c_{x_2} = 2 \), and the phases are guessed correctly.

In this setup, we are running a VP-detector with \( T \) steps on both modes containing the signal. The clicks can occur at steps \( t_1 \) and \( t_2 \), where we assume \( t_1 < t_2 \leq T \). Thus, the total probability can be expressed as the sum over all possible events. Here, an event is defined as the specific scenario in which clicks occur at steps \( t_1 \) and \( t_2 \), followed by correct phase guessing.

The factor \( \left(1 - e^{\frac{-N\mathcal{E}/2}{T}}\right)^2 e^{\frac{-(t_1 + t_2 - 2)N\mathcal{E}/2}{T}} \) represents the probability of obtaining a click exactly at step $t_1$ on one mode and another click exactly at step $t_2$ on the other mode. The additional factor of \( 2 \) accounts for the symmetry in the setup, where either mode could produce the first click. Finally, the $P_{VP}$ term stands for the probability of guessing the phases correctly with the remaining signal energy.

Instead, the probability of the VP detector not clicking in the last stage (Error 1) is
\begin{equation}
   P_{nc}(\mathcal{E},N)=\lim_{T\rightarrow\infty} \sum^{T}_{t_1=1}\sum_{t_2=t_1+1}^{T}2(1-e^{\frac{-N\mathcal{E}/2}{T}})^2e^{\frac{-(t_1+t_2-2)N\mathcal{E}/2}{T}} e^{\frac{-N(T-t_2)\mathcal{E}}{T}} \, ,
\end{equation}
obtained similarly to $P_S$ but substituting $P_{VP}$ with the probability of not clicking;{\color{red} upon this outcome, we make a random guess of the possible phase.} 
Finally, the probability of not getting any click at all (Error $2$) is $(1-e^{-N\mathcal{E}})$. In this case, differently from Error 1, we have gained no information about the message hence we make a random guess among all possible codewords. \\
\begin{figure}[t!]
    \centering
    \includegraphics[width=0.8\textwidth]{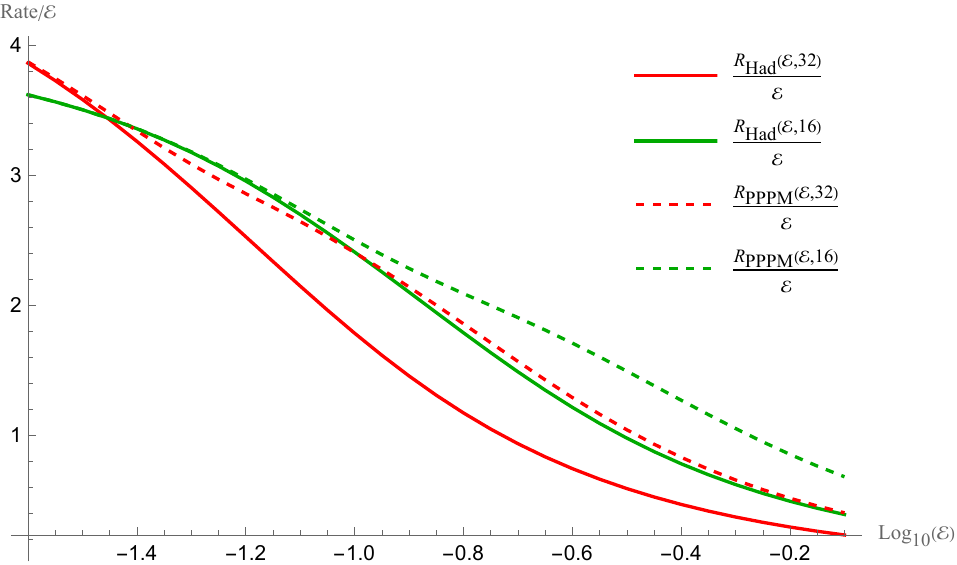}
    \caption{The comparison between the rate of generalized Hadamard code $R_{PPPM}$ in Eq.~\ref{gen.had} and the Hadamard code $R_{\text{Had}}$ in Eq.~\ref{Had rate}. }
    \label{fig:plot2}
\end{figure}
Once we have the conditional probabilities, doing some algebra we can write the mutual information and obtain the rate 
\begin{equation}
\begin{aligned}
   & R_{PPPM}(\mathcal{E},N,p) =\\
   &\frac{1}{N} \Bigg\{ 
        2N\Bigg( \, h_1 \Bigg[ 
            \frac{p \big( 1 - e^{-N\mathcal{E}} \big)}{2N} 
            + \frac{(1-p) \cdot 4 \binom{N}{2} \cdot \frac{2}{N} \cdot \frac{\big( 1 - e^{-\frac{N\mathcal{E}}{2}} \big) e^{-\frac{N\mathcal{E}}{2}}}{2}}{4 \binom{N}{2}} 
        \Bigg] 
        \\
        &- \frac{p}{2N} \Big[ 
            h_1 \big( P_{VP}(\sqrt{N\mathcal{E}} | \sqrt{N\mathcal{E}}) \big) 
            + h_1 \big( P_{VP}(\sqrt{N\mathcal{E}} | -\sqrt{N\mathcal{E}}) \big) 
        \Big]
        \\
        & - \frac{(1-p)}{4 \binom{N}{2}} \Bigg( 
            4 \binom{N}{2} \cdot \frac{2}{N} \cdot 
            h_1 \Big[ \frac{\big( 1 - e^{-\frac{N\mathcal{E}}{2}} \big) e^{-\frac{N\mathcal{E}}{2}}}{2} \Big] 
        \Bigg)\Bigg)
        \\
        & + 4 \binom{N}{2} \Bigg( h_1 \Bigg[ 
            \frac{1}{4 \binom{N}{2}} (1-p) 
            \big( P_{S}(\mathcal{E}, N) + P_{E}(\mathcal{E}, N) \big) 
        \Bigg]
        \\
        & - \frac{(1-p)}{4 \binom{N}{2}} \Big[ 
            h_1 \big[ P_{S}(\mathcal{E}, N) \big] 
            + h_1 \big[ P_{E}(\mathcal{E}, N) \big] 
        \Big]\Bigg)
        \\
        & + \binom{N}{2} \Bigg( 
            h_1 \Big[ \frac{4 (1-p) P_{\text{n.c.}}(\mathcal{E}, N)}{4 \binom{N}{2}} \Big] 
            - \frac{4 \binom{N}{2} (1-p) h_1[P_{\text{n.c.}}(\mathcal{E}, N)]}{4 \binom{N}{2}} 
        \Bigg) 
    \Bigg\}.
\end{aligned}
\end{equation}

\begin{figure}[t!]
    \centering
    \includegraphics[width=0.8\textwidth]{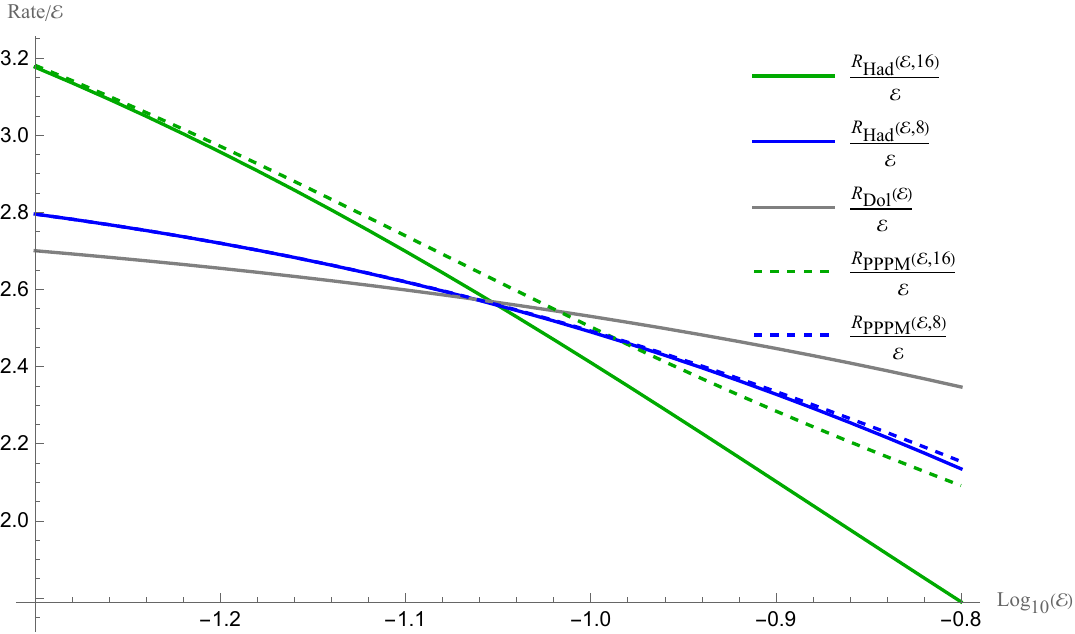}
    \caption{Comparison between the rate of the PPPM code $R_{PPPM}$ in Eq.~\ref{gen.had}, the Hadamard code $R_{\text{Had}}$ in Eq.~\ref{Had rate}, and the best single-symbol receiver, Dolinar in Eq.~\ref{Dol}.}
    \label{fig:plot3}
\end{figure}

Where \( p \) is the probability that messages are encoded into states with \( c = 1 \). As mentioned earlier, if \( p = 1 \), the code reduces to the Hadamard code. However, one can maximize over the parameter \( p \) to achieve a better rate compared to the Hadamard code. Therefore, we define the function
\begin{equation}\label{gen.had}
    R_{PPPM}(\mathcal{E}, N) := \max_p R_{PPPM}(\mathcal{E}, N, p).
\end{equation}
In Fig.~\ref{fig:plot2}, it can be observed that, at fixed $N$,  the PPPM code achieves a significant advantage over the Hadamard code, effectively extending the advantage of the latter to larger energies. Furthermore, we observe that the performance of PPPM at a given $N$ in this region moderately improves over a Hadamard code of sizze $N/2$. This suggests that a PPPM architecture is able to maintain performance across a wider energy region without changing $N$.

aAX In particular, if we focus on the transition region where the standard Hadamard code becomes sub-optimal with respect to single-symbol strategies such as Dolinar, i.e., $\mathcal E ~.1$, as shown in Fig.~\ref{fig:plot3}, we can observe that the PPPM code offers a moderate advantage over all previous strategies,  providing a bona fide JDR advantage in this energy range. 

\section{Discussion}
This work advances classical communication over lossy quantum channels by extending Hadamard codes to distribute signals across multiple modes. While standard Hadamard codes remain optimal in low-energy regimes, our approach demonstrates advantages as energy increases. While the overall performance gain is moderate, and the increased decoding complexity poses practical challenges, our PPPM code allows to use a single device of sufficiently large size across a wide energy range. Future research could explore expanding the code-space to include states where the signal is distributed across more than two modes with varying probabilities. 

\ack
F.K.~and M.R. acknowledge support from the project PNRR - Finanziato dall'Unione europea - MISSIONE 4 COMPONENTE 2 INVESTIMENTO 1.2 - ``Finanziamento di progetti presentati da giovani ricercatori'' - Id MSCA\_0000011-SQUID - CUP F83C22002390007 (Young Researchers) - Finanziato dall'Unione europea - NextGenerationEU.

\section*{References} 
\bibliographystyle{unsrt}
\bibliography{reference,library}


\end{document}